\documentclass[a4paper,11pt]{article}

\usepackage[top=2cm,bottom=2cm,left=2cm,right=2cm]{geometry}
\usepackage[pdftex]{graphicx} %per poter inserire le figure
\usepackage{amssymb,amsmath,amsthm,amsfonts}

\usepackage{graphicx} % Required for inserting images

\usepackage{xcolor}
\usepackage{xspace}
\usepackage{tabularx}
\usepackage{indentfirst}
\usepackage{subfigure}
\usepackage[small]{caption}
\usepackage{eucal}
\usepackage{eso-pic}
\usepackage{url}
\usepackage{booktabs}
\usepackage{afterpage}
\usepackage{parskip}
\usepackage{listings}
\usepackage{fancyhdr}
\usepackage{textcomp}
\usepackage{multirow}
\usepackage[utf8]{inputenc}   %per riuscire a scrivere gli accenti
\usepackage[english]{babel}   %per riuscire a scrivere gli accenti
\usepackage{setspace}
\usepackage{colortbl}
\usepackage{xcolor}   % per \textcolor e \definecolor
\usepackage[normalem]{ulem} % per \uline

\usepackage{subfiles}
\usepackage{biblatex}
\usepackage{hyperref}
\definecolor{BlueS}{RGB}{0, 0, 198}

\usepackage{authblk}

\title{Three-dimensional Lagrangian ecosystems: carbon dynamics and potential for artificial fertilization}

\author[1]{Stefano Campagnola}
\author[2]{Stephanie Dutkiewicz}
\author[2]{Michael J. Follows}
\author[1,*]{Enrico Ser-Giacomi}

\affil[1]{IFISC (CSIC-UIB), Institute for Cross-Disciplinary Physics and Complex Systems, Palma de Mallorca, Spain.}
\affil[2]{Department of Earth, Atmospheric and Planetary Sciences, Massachusetts Institute of Technology, 54-1514 MIT, Cambridge, MA 02139, USA.}
\affil[*]{enrico.sergiacomi@gmail.com}
\begin{document}

\maketitle
%\section{Running titles}
%\begin{enumerate}
%    \item Modeling carbon dynamics in Lagrangian oceanic ecosystems
 %   \item Carbon dynamics in three-dimensional Lagrangian ecosystems:  biophysical drivers of phytoplankton blooms and potential for artificial fertilization
  %  \item Carbon dynamics in Lagrangian ecosystems: identifying phytoplankton bloom drivers and assessing fertilization outcomes
   % \item Carbon and phytoplankton bloom dynamics in a Lagrangian Framework
   % \item Carbon and phytoplankton bloom dynamics in a Lagrangian Framework: Tracking artificial nutrient fertilization
%\end{enumerate}

\section{Abstract}

Transient supplies of nutrients to the surface ocean, both natural and artificial, stimulate blooms of phytoplankton and the formation of organic matter, driving air-sea gradients and uptake of CO$_2$. However, quantifying the associated carbon budget remains challenging, as it requires tracking the coupled biophysical evolution of water masses while they are transported, stretched and diluted. Here, we present an idealized three-dimensional model that describes biomass production and carbon dynamics within a Lagrangian patch in the ocean. The framework reproduces observed biogeochemical patterns from an artificial fertilization experiment and provides integrated metrics for the local carbon budget. Utilizing large ensembles of simulations, we examine the sensitivity of patch-scale primary production and carbon uptake to biochemical and physical factors. Our results show that patch dilution can enhance the ecosystem response, and how carbon uptake is sensitive to initial injected area, horizontal divergence and vertical diffusivity. In the context of renewed interest in ocean fertilization strategies for climate mitigation, our approach can thus provide quantitative tools to assess their efficacy and  potential.
%We argue that further investigation is required to constrain how biophysical drivers modulate the response to localized nutrient inputs, a prerequisite for robustly assessing both the efficacy and risks of ocean fertilization strategies.

\section{Introduction}

The Earth's carbon cycle involves exchanges among the atmosphere, ocean, and terrestrial reservoirs \cite{10.1007/978-3-642-84608-3_10}. The balance between these interconnected systems regulates the concentration of atmospheric carbon dioxide ($CO_2$), a major greenhouse gas driving climate change \cite{gmd-15-9075-2022, IPCC_2022_Ocean}. The gradient in the partial pressure of carbon dioxide ($pCO_2$) across the air-sea interface,  modulated by solubility \cite{WEISS1974203}, drives the exchange of $CO_2$ between the atmosphere and the ocean \cite{TAKAHASHI2009554}. 
A key mechanism that promotes carbon flux from the atmosphere into the ocean is the biological pump driven by photosynthetic organisms  (phytoplankton) which take up $CO_2$ in the surface sunlit layers, and export a fraction of this carbon to depth \cite{falkowski1998biogeochemical}.

Sudden increases in the availability of limiting nutrients can stimulate blooms: rapid and locally intensive growth in planktonic ecosystems \cite{bristow2017nutrients},  enhancing carbon absorption \cite{buesseler1998decoupling}. These phenomena have motivated growing interest in the artificial stimulation of phytoplankton blooms through the addition of key limiting nutrients. Ocean fertilization has been proposed as a potential approach to increase oceanic carbon sequestration as a carbon dioxide removal strategy \cite{martin1990glacial}. Consequently, substantial experimental and modelling efforts have sought to quantify its effectiveness and associated ecological impacts  \cite{debaar2005synthesis, boyd2007mesoscale}.

However, quantifying the associated carbon budget during such events remains challenging \cite{smetacek2008next}. Indeed, explicitly assessing bloom dynamics at the local scale requires tracking the biophysical evolution of the associated water mass over time from a Lagrangian perspective \cite{lehahn2017dispersion,ser2023lagrangian}. From an observational standpoint, this is difficult because of the sparse and non-synoptic nature of available measurements \cite{ioccg2015phytoplankton}. In addition, most current models of carbon uptake in fertilization experiments assume homogeneous conditions within the evolving water patch \cite{hannon2001modeling, fasham2006development}. However, recent work highlighted the important role of physical dilution \cite{lehahn2017dispersion}, horizontal advection, and the resulting spatial heterogeneity \cite{ser2023lagrangian} within the patch, which have important quantitative consequences for its integrated productivity. These findings emphasize the need for more realistic Lagrangian coupled biogeochemical–physical models to better quantify the drivers of local carbon dynamics during blooms and to assess the potential and risks of artificial fertilization as a carbon dioxide removal strategy.

Here we present an idealized, three-dimensional, Lagrangian model of a surface mixed-layer water parcel with associated plankton, nutrient and carbon dynamics. Lateral and vertical entrainment, as well as within-patch inhomegeneity, are parameterized. The model is a substantial modification of the framework developed in \cite{ser2023lagrangian} which adds the vertical dimension as well as biotic and abiotic carbon dynamics. Simulations of the Southern Ocean Iron Enrichment Experiment (SOIREE, \cite{boyd2000mesoscale}) predict biogeochemical evolution consistent with observations \cite{abraham2000importance, boyd2001southern}. Exploiting the efficiency of the Lagrangian model we then use large ensembles of simulations to quantify the sensitivity of integrated patch productivity and air-sea carbon flux  to initial patch sizes and specifics of the physical environment, including rates of dilution, upwelling and vertical mixing.
%and utilize an synthetic Lagrangian ensembles, to assess how dilution effects impact carbon uptake and how dynamical scales and concentration gradients set the optimal patch size for effective fertilization.  Finally, we quantify the contribution of vertical physical dynamics in modulating the overall response of the ecosystem to nutrient injections and resulting response on carbon uptake.

%%%% 400 words %%%%

\section{Results}

%\subsection{Three-dimensional Lagrangian ecosystems dynamics (opt 1)}
%\subsection{Tracer dynamics in a three-dimensional Lagrangian patch (opt 2)}
\subsection{Biochemical dynamics in three-dimensional Lagrangian ecosystems}

A  Lagrangian ecosystem in the ocean can be defined as a material patch of water along with its drifting chemical  and biological tracers.  Due to the effect of oceanic currents, the patch is deformed and diluted with time, affecting biochemical processes within the associated ecosystem  \cite{ser2023lagrangian}. In the following we develop an efficient, three-dimensional model to describe the explicit evolution of  such Lagrangian ecosystems including both physical and biochemical dynamics.

\begin{figure}
    \centering
    \includegraphics[width=0.9\linewidth]{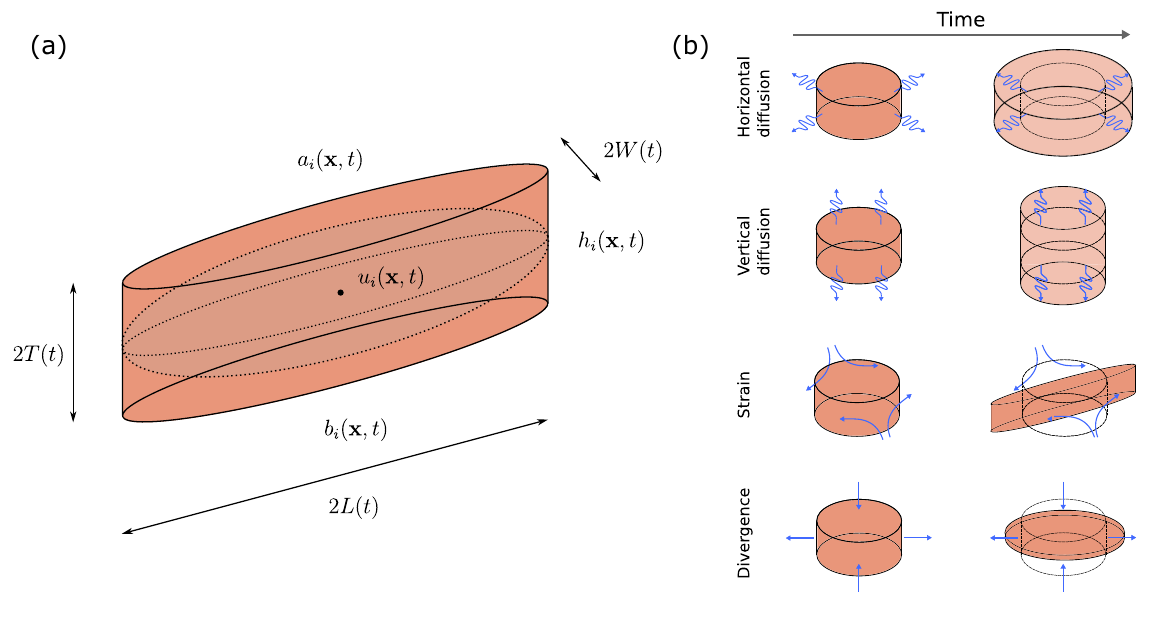}
   Three-dimensional Lagrangian patch dynamics. \caption{(a) Snapshot at time $t$ of a three-dimensional ellipsoidal Lagrangian patch of width $W(t)$, length $L(t)$ and thickness $T(t)$.  For a given tracer $i$, its distribution inside the patch is denoted as $u_i(\mathbf{x}, t)$ while the surrounding distributions around, above and below are indicated as $h_i(\mathbf{x}, t)$, $b_i(\mathbf{x}, t)$ and $a_i(\mathbf{x}, t)$, respectively. (b) Effects of horizontal and vertical diffusion, strain and divergence on a Lagrangian patch.  Diffusion in both directions increase the volume of the patch by diluting its content, while strain and divergence stretch the patch across the horizontal and the vertical, respectively.}
    \label{fig:patch_aggregated}
\end{figure}

A Lagrangian patch can be approximated by an ellipsoid of width $W(t)$, length $L(t)$ and thickness $T(t)$ (Fig. \ref{fig:patch_aggregated}a, Methods). Horizontal and vertical characteristic sizes are  $S^h(t) = W(t) + L(t)$ and $S^v(t) = 2T(t)$, while its horizontal projected area and volume are $A(t) = \pi W(t) L(t)$ and $V(t) = \frac{4}{3}\pi W(t) L(t) T(t)$. We identify four main physical mechanisms by which a  three-dimensional water patch changes its size and shape: horizontal strain,  divergence, horizontal diffusion and vertical diffusion \cite{ledwell1998mixing, williams2011ocean}. The horizontal strain rate $\gamma(t)$ measures the tendency of the patch to elongate along a given horizontal direction, while the divergence rate $\delta(t)$ accounts for patch flattening and radial expansion. Diffusivity rates, $\kappa^h(t)$ and $\kappa^v(t)$  represent the effect of unresolved scales of motion smaller than the characteristic patch sizes across the the horizontal and vertical direction, respectively (see Fig. \ref{fig:patch_aggregated}b). We also dynamically resolve the dependence of horizontal stirring and diffusion on  spatial scales so that $\gamma(t) = \gamma(S^h(t))$ and $\kappa^h(t) = \kappa^h(S^h(t))$ \cite{okubo1971oceanic, corrado2017general}.

Solving a three-dimensional advection-diffusion equation accounting for the above processes, we can predict how $W(t)$, $L(t)$ and $T(t)$ will evolve in time, and how this is reflected in the patch volume increase rate (Methods):

\begin{equation}
\begin{aligned}
\frac{d V(t)}{d t} & = \frac{4 \pi}{3}\left[\kappa^h(t)\left(\frac{W^2(t)+L^2(t)}{W(t) L(t)}\right) T(t)+\kappa^v(t) \frac{W(t) L(t)}{T(t)}\right] 
\end{aligned}
\label{eq:V-rate}
\end{equation} 

We first find that the effect of $\kappa^h(t)$ on patch volume expansion is regulated by the perimeter-to-area ratio times the thickness of the patch, while the contribution of $\kappa^v(t)$ is modulated by the area-to-thickness ratio. 
%\textcolor{red}{Note that,  in Eq. (\ref{eq:V-rate}), the influence of $\gamma(t)$ and $\delta(t)$ is not explicit, but it is embedded in the evolution of $W(t)$, $L(t)$, and $T(t)$ (Methods).}
%\textcolor{red}{Higher strain will implicitly promote a higher perimeter-to-area ratio, while high divergence will increase the area-to-thickness ratio.}
The relative rate of increase of the patch volume is defined as $\Omega(t) = \frac{1}{V(t)} \frac{dV(t)}{dt}$ -  referred to hereafter as the entrainment rate.

We then consider the concentration distributions of generic tracers inside the patch, as well as around, below and above (see Fig. \ref{fig:patch_aggregated}a). 
The three main processes that drive changes of tracer distributions within the Lagrangian patch are \cite{ser2023lagrangian}: (i) entrainment of surrounding water with potentially different concentrations; (ii) physical mixing within the patch interior; (iii) biochemical interactions between tracers (such as the uptake of carbon). Resolving these processes leads to a general master equation  governing the temporal evolution of every tracer inside the patch (Methods). To address spatial heterogeneity within the ecosystem explicitly and to solve the master equation, we apply Reynolds decomposition, thereby obtaining solutions for the first (spatial means) and the second moments (spatial variances and covariances) of the tracer concentrations (Methods, Eq. (\ref{eq:stat_model})).

We model community phytoplankton dynamics in the patch following r\cite{ser2023lagrangian} . Here we introduce an additional tracer, the difference in $CO_2$ partial pressure between the interior of the patch and the overlying atmosphere, $\Delta p(\mathbf{x}, t)$. Starting from the relation between $\Delta p(\mathbf{x}, t)$ and the distribution of Dissolved Inorganic Carbon ($DIC$), we can model the carbon dynamics to a good approximation as follows (see Methods):

\begin{equation}
\begin{aligned}
\left[ \frac{d \Delta p(\mathbf{x}, t)}{dt} \right]_{bio} & =   - \phi \frac{K_o K_g}{h} \Delta p(\mathbf{x}, t) + \frac{\phi}{\rho} \frac{1}{R_{N:C}} S_{upt}(\mathbf{x}, t) - \phi \Sigma_{\mathcal{T}} \frac{d \mathcal{T}(\mathbf{x}, t)}{d t}\\ 
\end{aligned} \label{eq:model_pco2}
\end{equation}

where the three terms on the right respectively reflect the impact of air-sea $CO_2$ exchange, biological uptake of carbon and  solubility forcing related to sea surface temperature change. 
%Further calculations allow us to extract from Eqs. (\ref{eq:model_bio}) and (\ref{eq:model_pco2}) the evolution of spatial means ($\langle N \rangle_{pat}$, $\langle P \rangle_{pat}$, $\langle \Delta p \rangle_{pat}$) as well as variances and covariances ($\langle N'^2 \rangle_{pat}$, $\langle P'^2 \rangle_{pat}$, $\langle \Delta p'^2 \rangle_{pat}$, $\langle N'P' \rangle_{pat}$, $\langle N'\Delta p' \rangle_{pat}$, $\langle P'\Delta p' \rangle_{pat}$) of the tracers' distributions  (Methods). 

We explicitly assess effect of nutrient injections on the patch-integrated air-sea carbon flux and primary production. To this end, we define the Lagrangian Carbon Flux (LCF) as the total mass anomaly (\textit{tons}) of carbon absorbed across the evolving patch surface over a time window of $\tau$ days:
\begin{equation}
LCF = - \int_{0}^{\tau} \rho K_o K_g \left( \langle \Delta p(t) \rangle_{pat} - \langle \Delta p^h(t) \rangle_{suh}\right) A(t) dt \label{eq:aic}
\end{equation}
where the term inside the integral represents the difference between the mean flux of $CO_2$ entering the patch at time $t$ and that in the horizontal surrounding waters. Similarly, we estimate the Lagrangian Primary Production (LPP) computed over the patch volume as:
\begin{equation}
LPP = - \int_{0}^{\tau}  \frac{1}{R_{N:C}} \left( \langle S_{upt}(t) \rangle_{pat} - \langle S^h_{upt}(t) \rangle_{suh} \right) V(t) dt \label{eq:aic}
\end{equation}
where we integrate the anomaly of  the biological term $S_{upt}$ relative to the surrounding waters over a time window of $\tau$ days.

%%% 700 words %%%% 

%\subsection{Example application: Fertilization experiments (opt 1)}
\subsection{A Lagrangian view on fertilization experiments}
%\subsection{A Lagrangian perspective of a fertilization experiment (opt 3)}

SOIREE was the first iron fertilization experiment conducted in the polar Southern Ocean \cite{boyd2000mesoscale, abraham2000importance}. The fertilized patch was tracked in situ for approximately two weeks after iron injection and by satellite over the following 40 days. Its Lagrangian design, based on the release of an inert tracer ($SF_6$), together with repeated measurements, makes SOIREE an ideal case study for assessing the local response of plankton ecosystems to fertilization \cite{bakker2001southern}. Here, we compare predictions from our Lagrangian model with in situ observations \cite{abraham2000importance, hannon2001modeling, bmodmopco2} and previous numerical simulations (SWAMCO, \cite{lancelot2000modeling, bmodmoSWAMCO}) over a 42-day period, focusing on the evolution of the fertilized patch and the associated iron, phytoplankton, and carbon concentrations (Methods).

%We simulate a $42$-day time window from 9 February, corresponding to the first iron infusion, until the last available SeaWiFS satellite observations. The modeled fertilized  patch has an initial diameter of 10 km \cite{bowie2001fate} and a thickness equal to the local mixed-layer depth, approximately $65$ m \cite{abraham2000importance}. Physical model parameters such as strain rate, divergence, and diffusivities were measured during the SOIREE experiment \cite{abraham2000importance, hannon2001modeling} and scaled with size according to \cite{okubo1971oceanic, corrado2017general} (see Methods). 

%\textcolor{red}{Biochemical parameters and initial conditions for the tracers' concentrations are derived from literature or inferred from experimental data to mimic the environment where the iron release took place \cite{abraham2000importance}, \cite{hannon2001modeling}, \cite{frew2001macronutrient} (Methods).}

\begin{figure}
    \centering
    \includegraphics[width=0.9\linewidth]{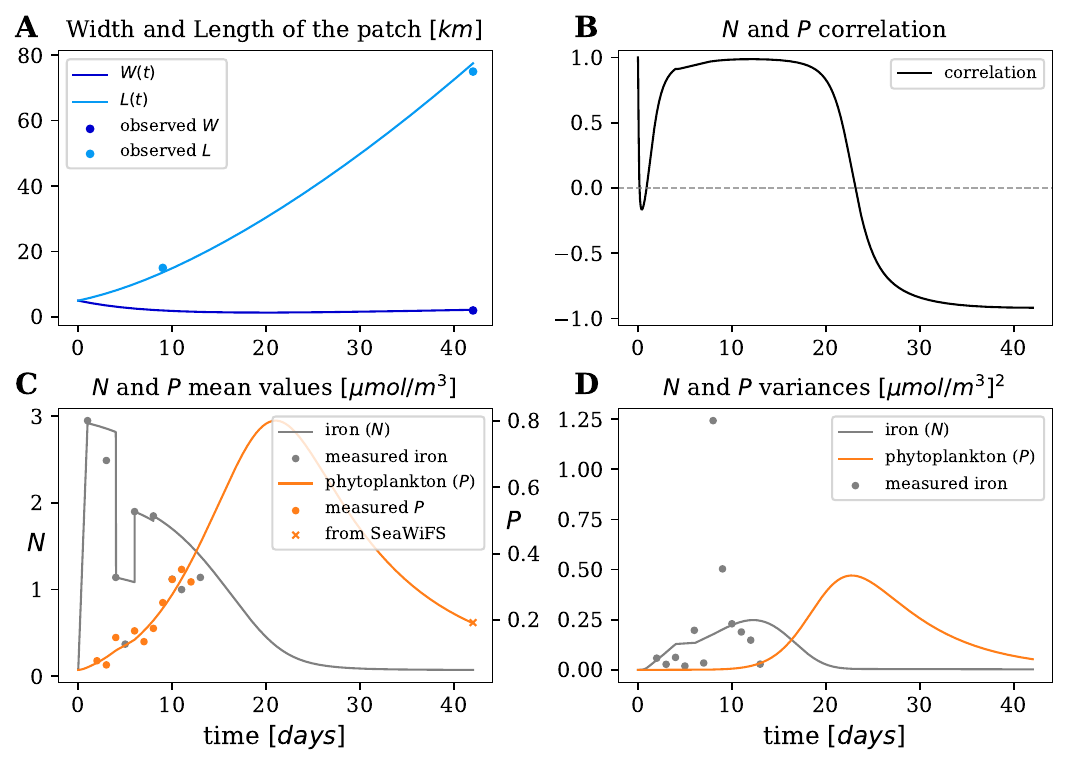}
    \caption{Simulation of the SOIREE fertilization experiment. Solid lines represent the time evolution of the model variables: width and length of the patch (panel A), iron and phytoplankton average concentration values (panel C), iron and phytoplankton concentration variances (panel D) and iron-phytolankton spatial correlation (panel B). In panel C we are adopting two vertical axes, the left one for iron concentration and the right one for phytoplankton concentration, due to their different scale of magnitude. Dots correspond to in-situ measurements, while crosses denote remote-sensed observations from satellites. The biomass of phytoplankton is expressed in iron currency.} 
    \label{fig:N-P_simple}
\end{figure}

\begin{figure}
    \centering
    \includegraphics[width=0.9\linewidth]{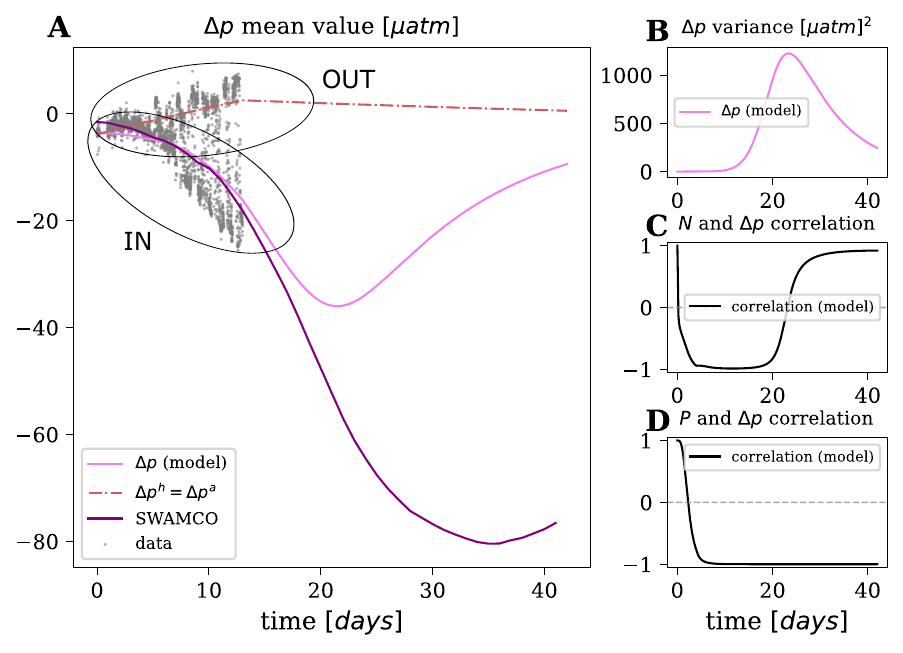}
    \caption{Simulated $pCO_2$ anomaly ($\Delta p$) and its relation with iron and phytoplankton during SOIREE. The gray small dots correspond to in-situ $\Delta p$ measurements. The lower cloud of points correspond to observations inside of the fertilized patch while the upper cloud represent $\Delta p$ measurements at the surrounding. Solid lines are the time evolution of models variables: model $\Delta p$ average concentration value against SWAMCO prediction (panel A), $\Delta p$ concentration variance (panel B), iron-$\Delta p$ spatial correlation (panel C) and phytoplankton-$\Delta p$ spatial correlation (panel D). Dash-dotted line in panel A represents modeled $\Delta p$ outside the patch, following the upper cloud of points.} 
    \label{fig:delta-p_simple}
\end{figure}

Our simulation reproduces the observed evolution of the SOIREE patch width $W(t)$ and length $L(t)$  (Fig. \ref{fig:N-P_simple}A). %We note that if we had instead used a constant entrainment rate (as is often assumed) there would have been an unrealistic exponential growth of the volume. During SOIREE, in situ casts reported only a modest decrease in the mixed layer depth. For this reason, and considering the low values of $\delta(t)$ and $\kappa^v(t)$ characteristic of the region, no significant change in thickness is obtained.
The initially high iron concentration inside the patch generates a strong gradient with the surrounding waters, enhancing spatial heterogeneity as interior and exterior waters are mingled (Fig. \ref{fig:N-P_simple}C). As the fertilization spurs a phytoplankton bloom, biological uptake of iron increases, dissolved iron concentrations decrease and, after about 25 days, the mean iron concentration returns to values close to the background conditions.  The phytoplankton bloom peaks after about 21 days, with average concentrations reaching nearly 16 times those of the surrounding waters.   During the first and last part of the simulation, modeled phytoplankton concentrations are consistent with in-situ measurements \cite{bcodmochl} and SeaWiFS satellite observations \cite{abraham2000importance}.

Iron spatial variance stabilizes after about 4 days, but phytoplankton variance peaks after several weeks (Fig. \ref{fig:N-P_simple}D). The tracers spatial correlation indicates that, at the initial stage of the bloom, phytoplankton biomass is strongly associated with iron-rich spots (Fig. \ref{fig:N-P_simple}B). 
As mean iron concentrations decrease to background levels, the correlation turns negative, reflecting phytoplankton-driven depletion of localized iron sub-patches, consistent with resource over-exploitation dynamics \cite{abraham1998generation}.

Carbon sequestration within the water patch is directly linked to the dynamics of the $pCO_2$ anomaly, $\Delta p$. Similar to phytoplankton, the spatial variance of $\Delta p$ increases rapidly after day 15 (Fig. \ref{fig:delta-p_simple}B). The correlation between nutrient concentration and $\Delta p$ is initially negative but becomes positive during the final stage of the experiment, consistent with phytoplankton growth enhancing carbon uptake and thereby reducing the $pCO_2$ anomaly (Fig. \ref{fig:delta-p_simple}C). In contrast, the correlation between phytoplankton and $\Delta p$ remains strongly negative throughout the simulation, reflecting the linear dependence of the biological uptake term $S_{upt}$ on $P$ (Fig. \ref{fig:delta-p_simple}D).

When comparing our model predictions with those of SWAMCO \cite{lancelot2000modeling}, a fine-tuned biogeochemical model specific for SOIREE but neglecting water patch physics, both simulations underestimate the mean $\Delta p$ anomaly relative to the data collected during the first $13$ days inside the patch (bottom part of the gray cloud, Fig. \ref{fig:delta-p_simple}A).
However, especially after day $20$, the model predictions diverge with regard to the $\Delta p$ anomaly, with SWAMCO predicting a much stronger response. Sensitivity studies with our model suggest that this discrepancy is partially due to the constant entrainment rate $\Omega(t) \simeq 0.03\ \mathrm{day^{-1}}$  \cite{hannon2001modeling} which we suggest is unrealistic. The significantly more complex biological simulation of SWAMCO may also contribute. Since no in situ measurements of $\Delta p$ are available after the end of the cruise at day $14$, there is no empirical basis to determine which prediction is more reliable.

The patch integrated air-sea flux of carbon, LCF, computed for the first $13$ days of the Lagrangian simulation (Fig. \ref{fig:delta-p_simple}) suggests an uptake of $32$ \textit{tons} of carbon, which compares well to the in situ observations from SOIREE of $31$ \textit{tons} \cite{bakker2001southern}. In contrast, SWAMCO predicted $130$ \textit{tons} \cite{hannon2001modeling}. This discrepancy is probably due to a different way of estimating the air-sea $CO_2$ flux.
The LCF from our model after 42 days is $690$ \textit{tons}: Only $\sim 5\%$ of it is absorbed during the first two weeks, reinforcing the point that in order quantitatively to assess the overall impact of fertilization,  ground-truthing in situ observations would need to extend beyond fifteen days \cite{yoon2018reviews}. A sensitivity experiment that neglects the patch dynamics and assumes a constant entrainment rate (as in SWAMCO) leads to an estimate of LCF of about $60\%$ higher.

When considering the impact of fertilization on primary productivity, we estimate a LPP of about $420$ \textit{tons} during the first $13$ days, comparable to the $400$ \textit{tons} of increased algal carbon measured by \cite{boyd2000mesoscale} following the $SF_6$ control volume. Other evaluations of the primary production enhancement by SOIREE are much higher, $1200$-$1400$ \textit{tons} \cite{hannon2001modeling, bakker2001southern}, but these studies probably consider a larger volume of integration. Our model estimates  $1640$ \textit{tons} of primary production inside the water patch after 42 days. The total organic carbon produced in the first 2 weeks is about $25 \%$ of this final amount, suggesting that primary production in the Lagrangian ecosystem reacts faster than the carbon flux to a nutrient pulse, as proposed by \cite{de2005synthesis, buesseler2008ocean}.

%Finally, we assess how assuming a constant entrainment rate in the simulation ($\Omega(t) \simeq 0.03\ \mathrm{day^{-1}}$) can bias the estimated fertilization outcome. Under this assumption, the simulated $pCO_2$ levels (Fig. \ref{fig:delta-p_simple}A, dotted line) exhibit a larger decrease than in the main simulation, approaching the values predicted by the SWAMCO model. \commentos{mixing the two phrases}As a direct consequence, the LCF is overestimated, reaching 1080 \textit{tons} after 42 days of integration.Neglecting the patch dynamics and the resulting time-dependent entrainment rate therefore leads to an increase in the estimated LCF of about $60 \%$.

%%%% 1070 words

%\subsection{Response to nutrient injection for ensembles of Lagrangian ecosystems (opt 1)} 
%\subsection{Ensembles of Lagrangian ecosystems analysis (opt 2)}
\subsection{Nutrient injections across ensembles of Lagrangian ecosystems}

In this section, enabled by the efficiency of the model-structure, we build large ensembles of simulations (around 3000) to investigate how water patches subjected to different physical forcings or characterized by different initial sizes respond to nutrient injection. 
%in which each synthetic ecosystem is subject to diverse physical forcing conditions while keeping  the amount of nutrient injected fixed. 
%What we want to assess is how much time-dependent entrainment rates, initial patch size and spatial heterogeneity  influence the outcome of the nutrient injection in terms of integrated quantities.
%This allow to highlight the main drivers of ecosystem carbon dynamic.
We first examine how horizontal diffusivity, strain and initial patch area can affect development of the bloom, also comparing an infinitely well-mixed scenario against our realistic heterogeneous approach. 
Secondly, vertical diffusivity and divergence are included in a full heteroegeneous setup and analyzed together with horizontal dynamics. 
We use the dilution factor -- defined as the ratio between the average patch volume during the simulation and the initial one -- as an appropriate metric to represent the global evolution of the single Lagrangian ecosystems.

\begin{figure}
    \centering
    \includegraphics[width=0.9\linewidth]{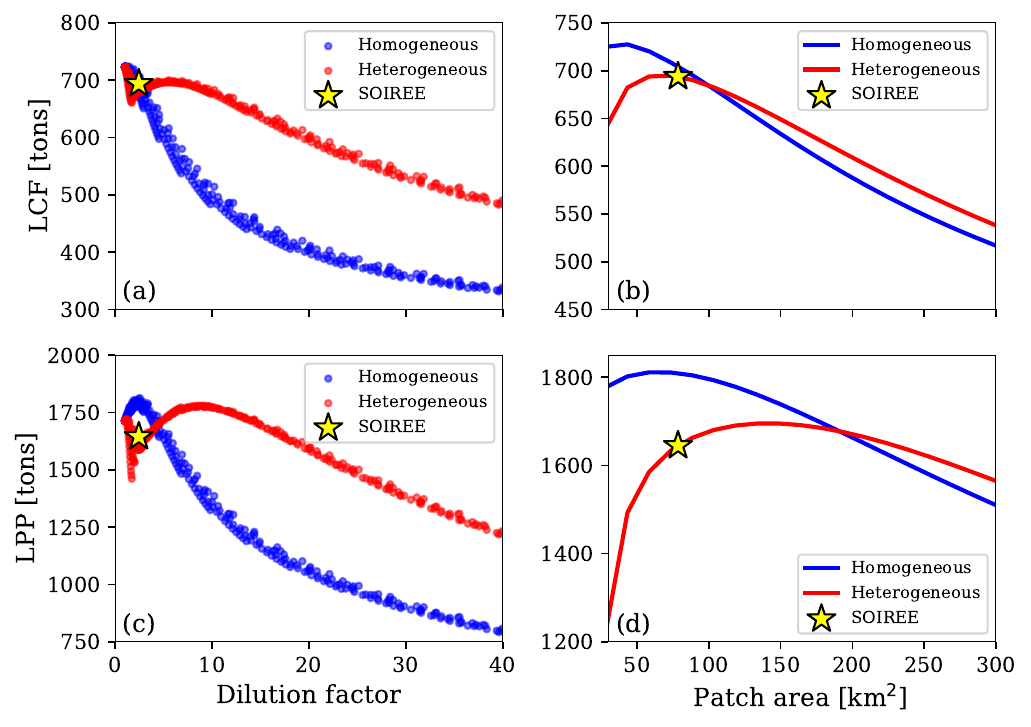}
    \caption{Lagrangian Carbon Flux (LCF) and Lagrangian Primary Production (LPP) after 42 days across different values of patch dilution factor and initial patch injected area. Red and blue dots/lines refer to model configurations where second moments of tracer concentrations are resolved or neglected, respectively. The yellow star refers to the SOIREE simulation.} 
    \label{fig:dil_factor_and_init_patch}
\end{figure}

We first consider the LCF and LPP obtained by different combinations of  horizontal diffusivity and strain $(\kappa^h_0,\ \gamma_0)$ while keeping the vertical diffusivity and divergence $(\kappa^v_0,\ \delta_0)$ fixed (Fig. \ref{fig:dil_factor_and_init_patch}a-\ref{fig:dil_factor_and_init_patch}c). 
In a setup assuming a homogeneous patch interior (i.e. neglecting tracers second moments), LCF always decreases as the dilution gets stronger, meaning that any intrusion of surrounding water leads to a lower uptake of $CO_2$ from the atmosphere. However, LPP benefits from small dilutions even in a well-mixed environment, probably due to a wider persistence of phytoplankton biomass peaks.  
In the more realistic scenario representing within patch heterogeneity of tracers (i.e. resolving second moments) we observe significant changes in sensitivities for both LCF and LPP. 
Overall, consistently with previous findings \cite{ser2023lagrangian}, spatial heterogeneity increases the LPP in response to fertilization with respect to the simplified well-mixed assumption.
This change is related to the contribution of the second order terms and the positive covariance on phytoplankton growth.
Moreover, we note that while LPP is maximized for medium dilutions, LCF is optimal at low dilution rates. 
The latter is mainly due to the differences in equations \ref{eq:model_bio}, \ref{eq:model_pco2_3} and the role of the air-sea $CO_2$ flux term.

% combo of different aspects, the stronger one seems that dp dynamic is retarded in time with respect to S_upt (based on the product N*P, dp e P go in parallel, but N goes to zero quickly). So maybe is the reason we stop at 42 days
% another aspect could be the presence of the flux_term in the equation of dp which fasten the reequilibration near the end of the simulation

We explore the influence of the horizontal spatial scale of fertilization on the ecosystem response.
Previous studies on passive tracers \cite{xiu2010modeling} also highlighted the importance of the initial patch size and how this can affect the rate of dispersal. 
Given our approach, this can be modeled by infusing the same absolute nutrient load over different initial patch areas. 
We expect, depending on the initial injection scale, changes in the resource concentration and gradient as well as in the physics experienced by the ecosystem (Figures \ref{fig:dil_factor_and_init_patch}b-\ref{fig:dil_factor_and_init_patch}d). %\commentos{see if there are articles talking about that also regarding natural deposition of iron, they say only that are very big patches}
%Figures \ref{fig:dil_factor_and_init_patch}b-\ref{fig:dil_factor_and_init_patch}d show the LCF and LPP that can be obtained by fertilizing different portions of the ocean across an interval of sizes.
Here, again, the homogeneous and heterogeneous setups show significative differences regarding the optimal intervals of initial patch area.
A well-mixed patch is generally more productive if the nutrient is concentrated in region smaller than $70\ km^2$, while if we account for heterogeneity fertilizing larger patches  (from $70$ to $150\ km^2$) results in stronger ecosystem responses. 
Interestingly, the real SOIREE initial patch size (yellow star in the figure) turns out to be of optimal size for carbon dioxide capture in a heterogeneous patch, and even the LPP is close its maximum potential. 
The decrease of LCF and LPP  for large patches is related to a slower patch volume increase due to strain and diffusivity scaling laws, while the non linear plankton dynamics can explain the reduction in LCF and LPP  observed for  small initial patch size suggesting a saturated growth condition.

\begin{figure*}  % figura in doppia colonna
    \centering
    \includegraphics[width=0.9\textwidth]{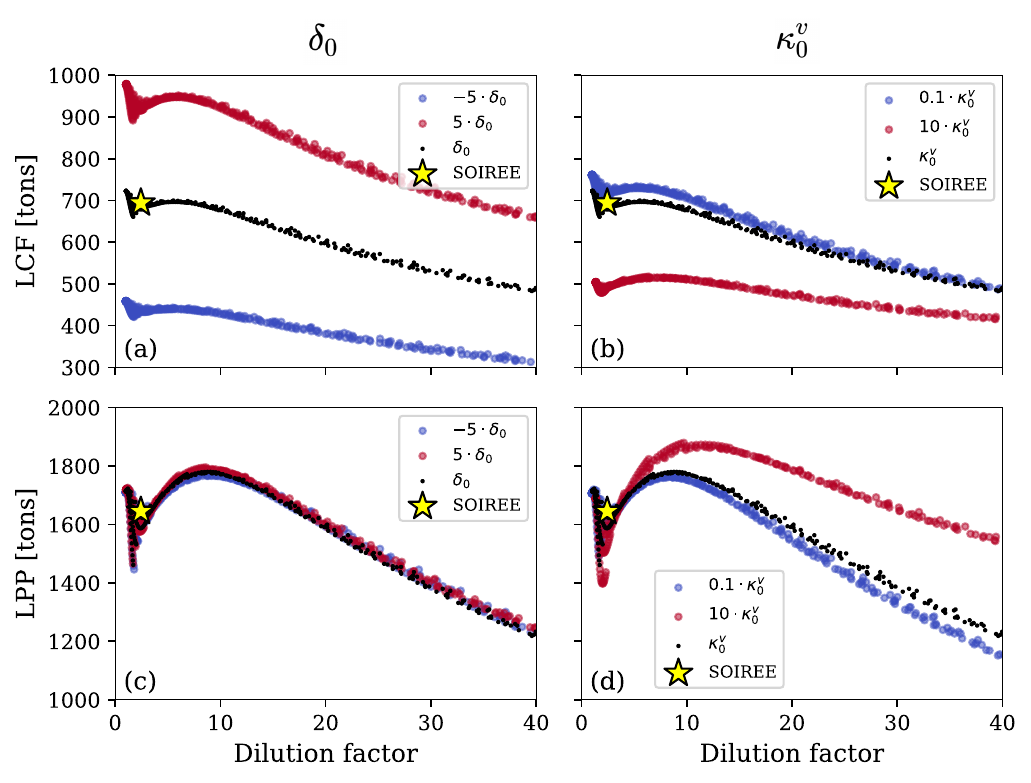}
    \caption{Lagrangian Carbon Flux (LCF) and Lagrangian Primary Production (LPP) after 42 days across different values of dilution factors and with different vertical dynamics. All panels refers to a heterogeneous model setups. In black we report the red curves of Fig. \ref{fig:dil_factor_and_init_patch}a-\ref{fig:dil_factor_and_init_patch}c, while in dark red and cyan we plot simulations outputs varying the initial horizontal divergence $\delta_0$ and vertical diffusivity $\kappa_0^v$. The yellow star refers to the SOIREE simulation.} 
    \label{fig:dilution_factor_vertical}
\end{figure*}

The last aspect of physical forcing that we address is the role of the vertical dynamic while keeping constant surrounding concentrations.
%Again, we explore different combinations of $(\kappa^h_0,\ \gamma_0)$ that give rise to different dilution factors.
In the first ensemble of simulations, we fix the value of the vertical diffusivity $\kappa^v_0$  while varying the divergence $\delta_0$ in magnitude and sign, and explore the impact of different dilutions. 
We find that a positive divergence leads to a larger patch horizontal area while reducing the overall patch volume. 
This geometric effect reduces the homogenization rate, inducing higher spatial variances and covariances.
As a consequence, phytoplankton concentrations increase relative to the baseline case.
These combined effects explain the observed behavior of the LCF, which is primarily area-driven. In particular, LCF increases by approximately 40\% across all horizontal dilution conditions (Fig. \ref{fig:dilution_factor_vertical}a).
In contrast, the LPP (Fig. \ref{fig:dilution_factor_vertical}c), which is sensitive to volumetric effects, responds differently: the reduction in volume counterbalances the increase in phytoplankton concentration, resulting in negligible net changes.

Finally we focus on exploring different levels of vertical diffusivity $\kappa^v_0$ keeping the divergence $\delta_0$ fixed. 
The results show a lower LCF and higher LPP for increasing vertical diffusivity (see Fig. \ref{fig:dilution_factor_vertical}b-\ref{fig:dilution_factor_vertical}d)   
with variations  ranging from $5 \%$ to $30 \%$ with respect to the baseline (black line). 
The decrease in LCF is probably caused by higher exchange of carbon between the water patch and deeper layer, replacing part of the atmosphere-to-sea flux from above.
On the other hand, the increase of LPP is attributable to a stronger vertical entrainment rate fostering heterogeneity, coupled with a higher exchange of iron with the bottom layer.
All in all, our results suggest that positive divergence is always beneficial, while smaller or higher values of vertical diffusivity can enhance contrasting aspects of the bloom response. 
We also stress that our model does not resolve modulation of the growth rate as function of the patch average depth. 
We speculate that a thickness-dependent growth rate would enhance the effect of positive divergence exploiting the process of patch thinning.

%Redfield-like ratios in our model represent community-mean requirements for iron and carbon-to-chlorophyll relationships. Exploring realistic Southern Ocean ranges shows that variability in $R_{Fe:C}$ and $R_{C:Chl}$ strongly affects model outputs, with LCF and LPP varying by factors of 3–8 over the simulation.

%%%% approx. 900 words

\section{Conclusions}

In this work, developing from \cite{ser2023lagrangian}, we introduced a novel theoretical framework  to describe the three-dimensional bio-physical dynamics of Lagrangian ecosystems embedded in evolving water patches. We explicitly resolve processes including local upwelling and vertical diffusivity, as well as the dynamics of  the $CO_2$ partial pressure anomaly to provide estimates of air-sea carbon uptake.

Our approach reproduces the SOIREE observations realistically, capturing both the physical and biological evolution of the bloom, estimating a $pCO_2$ anomaly trend which follows the initial in-situ measurements \cite{abraham2000importance,boyd2001southern}. Direct observationsof the late stage of such blooms, potentially using Lagrangian autonomous platforms \cite{claustre2020observing}, would provide very valuable constraints as model results suggest that most biomass growth occurs during this period.  A fundamental challenge is also the identification and tracking of the fertilized water patch, as uncertainties in its total volume remain a major source of discrepancies in estimates of integrated quantities reported in the literature.

%Several studies have discussed the applicability of Lagrangian frameworks in both biogeochemical and oceanographic models \cite{villa2020ocean, abraham2000importance, kranz2020lagrangian}. However, only recently has a connection between a Lagrangian perspective and ecosystem heterogeneity been established \cite{ser2023lagrangian}. In this work, we extend the results of \cite{ser2023lagrangian} by analyzing such implications in terms of $pCO_2$ reduction and atmospheric carbon uptake, explicitly accounting for physical dilution and mixing processes.

%In this work we introduced a novel, fully three-dimensional Lagrangian ecosystem model that allowed us to account for processes involving vertical exchanges, such as upwelling and enhanced vertical diffusivity.  

%From a biochemical point of view, chemical ratios emerge as the most critical parameters in our model for the estimation of $CO_2$ draw down and primary production throughout the bloom expansion (see Fig. \ref{fig:ratios}). The variability we observe is a reflection of the possible outcomes that a fertilization experiment can attain in environments with different conditions and ecological assemblages. 

Our ensembles of simulations highlight the key role of dilution in generating tracer heterogeneity that can in turn enhance the biomass growth of planktonic ecosystems \cite{lehahn2017dispersion, ser2023lagrangian}. These findings have important implications for the design of fertilization experiments, as they suggest that coherent and stable regions of the ocean may not necessarily represent the most favorable sites for maximizing community production \cite{yoon2018reviews}.

We also find that there is an optimal choice in the initial patch size  that maximizes the carbon flux and primary production. This is determined by the interplay between scaling laws of the physical drivers  and size-dependent concentration gradients.
%Finally, we discuss how the interplay between scaling laws in the ocean and high concentration gradients determines a favorable range of initial patch sizes to be fertilized, in order to maximize the outcome of $CO_2$ absorption and bloom productivity. 
Thus, an extensive analysis on realistic mesoscale and submesoscale flows \cite{xiu2010modeling} acting on different patch sizes of biological tracers could be of large interest for future fertilization studies.

Finally, we find that increased vertical upwelling from below the mixed layer and the associated horizontal divergence enhance net carbon uptake during the bloom. In contrast, the magnitude of vertical diffusivity has more complex effects on carbon flux and biomass production. We speculate that such vertical processes may play an even greater role at the submesoscale, where vertical motions become increasingly important \cite{rosso2016quantifying, zhu2024observations}.

%Extensions of our approach could also include a thickness-dependent light availability to modulate the average growth rate experienced by phytoplankton.
%In any case, we expect that these aspects would coherently amplify the effect of the divergent flow on the ecosystem production.

%Though the third (vertical) dimension appeared to play a relatively minor role during the SOIREE campaign, in other dynamical contexts (different dilutions, initial patch size) we showed that its contribution may become significant (see Fig. \ref{fig:dilution_factor_vertical}). 

%Moreover, model sensitivy to stoichiometry (see Supplementary) suggests to consider an environment in which chemical ratios change dynamically.
%Despite recent advances in dynamic quota, trait-based, and optimality-based phytoplankton models, substantial uncertainties remain \cite{flynn2003modelling, pahlow2009chain} and this is considered a major challenge for marine ecosystem description.

Overall, we demonstrate that explicitly resolving both the physical and biogeochemical evolution of fertilized water patches is essential for accurately quantifying bloom dynamics and carbon uptake. As interest in artificial nutrient injections as a potential climate change mitigation strategy continues to grow, such process-based frameworks will be instrumental for the design, interpretation, and assessment of future fertilization strategies.

%%% 540 words

\newpage

\section{Methods}

\subsection{Advection-diffusion equation and patch physical transformation}

%We extend the model introduced in \cite{enrico} to embody the third spatial dimension, which is the depth. 

A Lagrangian water patch can be described as a three-dimensional ellipsoid centered in $\mathbf{X}(t) = (x_0, y_0, z_0)$ which contains the majority of the volume occupied by an ideal passive tracer that is dispersing from the same initial point (Supplementaryt Fig. 1).
%\commentos{maybe ref here and put in supplementary the old image with the frame of reference}
% (of distribution $\theta (\mathbf{x},t)$)
The adopted moving frame of reference is also centered in $\mathbf{X}(t)$, translating and rotating horizontally with the patch (tilting is assumed to be negligible).
The semi-axes of the ellipsoid are supposed to be aligned with the horizontal and vertical directions, that we denote with $\mathbf{\hat{w}}(t)$, $\mathbf{\hat{l}}(t)$ and $\mathbf{\hat{t}}(t)$. 
%We can call their size $W(t)$, $L(t)$ and $T(t)$, in this way, we can identify $2 W(t)$ as the patch width, $2 L(t)$ as the patch length and $2 T(t)$ as the patch thickness.

Patch shape modification can be caused by horizontal stirring, divergence and diffusion. 
The first two can be modeled through the local velocity field $\mathbf{v}(\mathbf{x}, t)$ acting on the tracer patch, which is approximated, at first order, by a three-dimensional stagnation point flow:

\begin{equation}
    \mathbf{v}(\mathbf{x}, t) = \left(\left(-\gamma(t) + \frac{\delta(t)}{2}\right) x; \ \ \ \ \left(+\gamma(t) + \frac{\delta(t)}{2}\right) y ; \ \ \ \ - \delta(t) z   \right) \label{eq:velocity}
\end{equation}

Here we are assuming that horizontal stirring acts along the $y$ direction, which is aligned with the $\mathbf{\hat{l}}$ axis. 
The quantities $\gamma(t)$ and $\delta(t)$ represent time dependent horizontal strain and divergence rates respectively, in a way that vertical dimension dynamic is independent of the horizontal one. 

Diffusion contribution tends to enlarge the patch, diluting its content with the outside. 
We again distinct between time dependent horizontal and vertical diffusivity $\kappa^h(t)$ and $\kappa^v(t)$, since they are order of magnitude different \cite{hannon2001modeling}. 
The diffusion that we are treating here represents unresolved scales of the velocity field, typically much smaller than the patch characteristic size.

In the Lagrangian frame of reference, the effects of the aforementioned strain, divergence and diffusion can be joined into an advection-diffusion equation of the patch distribution $\theta(\mathbf{x}, t)$:

\begin{equation}
\frac{\partial \theta(\mathbf{x}, t)}{\partial t}+\mathbf{v}(\mathbf{x}, t) \cdot \mathbf{\nabla} \theta(\mathbf{x}, t)=\kappa^h(t)\left(\frac{\partial^2 \theta(\mathbf{x}, t)}{\partial x^2}+\frac{\partial^2 \theta(\mathbf{x}, t)}{\partial y^2}\right)+\kappa^v(t) \frac{\partial^2 \theta(\mathbf{x}, t)}{\partial z^2} \label{eq:adv_dif_gen}
\end{equation}

What we want to understand now is how the patch distribution geometrically evolves in time due to the presence of these "shaping" mechanism (strain, divergence and diffusion). 
The spatial standard deviations $\sigma_x$, $\sigma_y$ and $\sigma_z$ of the patch distribution $\theta(\mathbf{x}, t)$ represent a good proxy of the patch size. 
%Let's indeed find the spatial variances that can be computed as
Following the same approach of \cite{ser2023lagrangian}, we can retrieve how the respective variances are expected to evolve in time

\begin{equation}
\begin{aligned}
\frac{d \sigma_x^2(t)}{d t} & =+2 \kappa^h(t)+2\left(-\gamma(t)+\frac{\delta(t)}{2}\right) \sigma_x^2(t) \\
\frac{d \sigma_y^2(t)}{d t} & =+2 \kappa^h(t)+2\left(\gamma(t)+\frac{\delta(t)}{2}\right) \sigma_y^2(t) \\
\frac{d \sigma_z^2(t)}{d t} & =+2 \kappa^v(t)-2 \delta(t) \sigma_z^2(t)
\end{aligned} \label{eq:std}
\end{equation}

We can define the width, length and thickness of the water patch as multiples of the spatial standard deviations by a factor $n$, indeed $W(t) = n \sigma_x(t)$, $L(t) = n \sigma_y(t)$ and $T(t) = n \sigma_z(t)$. 
The choice of $n \geq 1$ is arbitrary, but is linked mathematically to the percentage of $\theta(\mathbf{x}, t)$ we are sure to include into the ellipsoid. 
The relations we find are:

\begin{equation}
\begin{aligned}
\frac{d W(t)}{d t} & =+ \frac{\kappa^h(t)}{W(t)}+\left(-\gamma(t)+\frac{\delta(t)}{2}\right) W(t) \\
\frac{d L(t)}{d t} & =+ \frac{\kappa^h(t)}{L(t)}+ \left(\gamma(t)+\frac{\delta(t)}{2}\right) L(t) \\
\frac{d T(t)}{d t} & =+ \frac{\kappa^v(t)}{T(t)} - \delta(t) T(t)
\end{aligned} \label{eq:WLT}
\end{equation}

In this way, equations (\ref{eq:WLT}) can be used, together with the derivation chain rule, to retrieve

\begin{equation}
\begin{aligned}
\frac{d V(t)}{d t} & = \frac{4 \pi}{3}\left[\kappa^h(t)\left(\frac{W^2(t)+L^2(t)}{W(t) L(t)}\right) T(t)+\kappa^v(t) \frac{W(t) L(t)}{T(t)}\right] \\ & = \left(\frac{d V(t)}{d t} \right)^h + \left(\frac{d V(t)}{d t} \right)^v
\end{aligned}
\label{eq:V-rate}
\end{equation} 

and 

\begin{equation}
\frac{d A(t)}{d t}=\pi \left[\kappa^h(t)\left(\frac{W^2(t)+L^2(t)}{W(t) L(t)}\right) +\delta(t) W(t) L(t) \right]
\end{equation}

Notice that the rate of change of $V(t)$ seems to depend only on the diffusive rates $\kappa^h(t)$ and $\kappa^v(t)$, but in reality there is a hidden dependence also on $\gamma(t)$ and $\delta(t)$ that, underneath, determine the values of $W(t)$, $L(t)$ and $T(t)$ through (\ref{eq:WLT}). 
This reflects the known impact of advection in enhancing the effective dispersal in physical systems of this kind \cite{n2010chemical}.

\subsection{Size-dependent laws for strain, divergence and diffusion}

Many experiments in the literature have shown how the magnitude of the horizontal stirring and diffusion effects depends on the spatial scale considered \cite{okubo1971oceanic, corrado2017general}.
In contrast, vertical dynamic, due to ocean stratification, is quite homogeneous inside the mixed layer, and so vertical diffusivity and divergence can be assumed to be constant in time \cite{ledwell1993evidence}. 
The temporal dependence of these deformation rates is set as:
%Hence, to remain general, we can connect the temporal dependence of $\gamma(t)$ and $\kappa^h(t)$ to patch characteristic sizes $S^h(t)$ and $S^v(t)$:

\begin{equation}
\begin{aligned}
\gamma(t) & = \gamma(S^h(t)) = \gamma_0 \left( \frac{S^h(t)}{S^h_0}\right)^{-\frac{2}{3}} \\ \kappa^h(t) & = \kappa^h(S^h(t)) = \kappa_0^h \left( \frac{S^h(t)}{S^h_0}\right) \\ 
\delta(t) & = \delta_0 \\
\kappa^v(t) & = \kappa^v_0 \\
\end{aligned} \label{eq:4param_scaling}
\end{equation}

where strain $\gamma(t)$ and horizontal diffusivity $\kappa^h(t)$ scalings are linked to the patch horizontal characteristic size $S^h(t) = W(t) + L(t)$. The scaling parameter $S^h_0$ represents instead at which scale we expect $\gamma(t) = \gamma_0$ and $\kappa^h(t) = \kappa^h_0$.

\subsection{Tracers evolution inside the patch}

We characterize biochemical and organism tracer dynamics within the Lagrangian ecosystem enclosed by the patch boundaries. We define the concentration distribution of a generic tracer $i$ inside the patch as $u_i(\mathbf{x}, t)$.  The changing volume of the patch is driven by the entrainment of water from the surrounding \cite{ser2023lagrangian}. As such, we also account for tracer's concentration around, below and above the patch as $h_i(\mathbf{x}, t)$,  $b_i(\mathbf{x}, t)$ and $a_i(\mathbf{x}, t)$, respectively (see Fig. \ref{fig:patch_aggregated}a).

Following the approach of \cite{ser2023lagrangian}, we identify three main processes that regulate the  tracer concentration distributions $u_i(\mathbf{x}, t)$ within a Lagrangian water patch:

\begin{itemize}
    \item \textit{entrainment} of external water with different tracer concentration: $\mathcal{E}[u_i(\mathbf{x}, t); h_i(\mathbf{x}, t), a_i(\mathbf{x}, t), b_i(\mathbf{x}, t)]$. It represents the net effect of dilution on the patch ecosystem.
    \item \textit{internal mixing} which tend to homogenize the tracer concentration $u_i(\mathbf{x}, t)$: $\mathcal{M}[u_i(\mathbf{x}, t)]$. It accounts for submesoscale processes happening inside the patch. 
    \item \textit{biogeochemical interaction} between different tracers: $\mathcal{I}[u_i(\mathbf{x}, t), \{ u(\mathbf{x}, t)\}]$. The term $ \{ u(\mathbf{x}, t)\}$ denotes the set of all tracers taken into account.
\end{itemize}

These three contributions combine to form a general master equation for $u_i(\mathbf{x}, t)$:

\begin{align}
\frac{du_i(\mathbf{x}, t)}{dt} & = \mathcal{E}[u_i(\mathbf{x}, t); h_i(\mathbf{x}, t), a_i(\mathbf{x}, t), b_i(\mathbf{x}, t)] + \mathcal{M}[u_i(\mathbf{x}, t)] + \mathcal{I}[u_i(\mathbf{x}, t), \{ u(\mathbf{x}, t)\}] \nonumber \\ & = \left[ \frac{du_i(\mathbf{x}, t)}{dt} \right]_{sur} + \left[ \frac{du_i(\mathbf{x}, t)}{dt} \right]_{mix} + \left[ \frac{du_i(\mathbf{x}, t)}{dt} \right]_{bio} \label{eq:master}   
\end{align}

where we have adopted a shorthand notation with $\textit{sur}$, $\textit{mix}$ and $\textit{bio}$ to denote the specific partition of the tracer derivative.

In order to simplify the mathematics and, at the same time, still describe a system with a certain amount of heterogeneity, we introduce properly some statistical quantities related to the tracer distributions. 
We make use of the Reynolds decomposition \cite{kundu2015fluid}, which express a given scalar field in terms of its average value and the fluctuations around it. 
In our case, the tracer distribution $u_i(\mathbf{x}, t)$ can be split into

\begin{equation}
u_i(\mathbf{x}, t) = \langle u_i \rangle_{pat} (t) + u_i'(\mathbf{x}, t) \label{eq:reynolds}
\end{equation}

where the mean value is defined as 

\begin{equation}
\langle u_i \rangle_{pat} (t) = \frac{1}{V(t)} \int_{pat} u_i(\mathbf{x}, t) dV \label{eq:u_mean}
\end{equation}

and $pat$ stands for the ellipsoidal patch.

The fluctuation term $u_i'(\mathbf{x}, t)$ can be used to determine variances and covariances, that are the second moments of the tracer distribution:

\begin{equation}
\langle u'_i u'_j \rangle_{pat} (t) = \frac{1}{V(t)} \int_{pat} u'_i (\mathbf{x}, t) u'_j (\mathbf{x}, t) dV \label{eq:u_cov}
\end{equation}

Around the water patch, we identify three different regions: water that is present around, above and below. 
Tracer concentrations in these locations are indeed described separately as $h_i(\mathbf{x}, t)$, $a_i(\mathbf{x}, t)$ and $b_i(\mathbf{x}, t)$, where $h$ stands for \textit{horizontal} surrounding, $a$ for \textit{above} and $b$ for \textit{below}. 
The definition of statistical quantities for them is straightforward, since they involve the same integrals of equations (\ref{eq:u_mean}), (\ref{eq:u_cov}), but in different regions of space (\textit{suh}, \textit{sua}, \textit{sub}).

The master equation can finally be rewritten as a system of scalar equations involving all the tracers taken into account and preserving the structure with $\textit{sur}$, $\textit{mix}$ and $\textit{bio}$ components:

\begin{equation}
\begin{aligned}
\frac{d \langle u_i \rangle_{pat}}{dt} & = \left[ \frac{d \langle u_i \rangle_{pat}}{dt} \right]_{sur} + \left[ \frac{d \langle u_i \rangle_{pat}}{dt} \right]_{bio} \\ \frac{d \langle u_i^{\prime 2} \rangle_{pat}}{dt} & = \left[ \frac{d \langle u_i^{\prime 2} \rangle_{pat}}{dt} \right]_{sur} + \left[ \frac{d \langle u_i^{\prime 2} \rangle_{pat}}{dt} \right]_{mix} + \left[ \frac{d \langle u_i^{\prime 2} \rangle_{pat}}{dt} \right]_{bio} \\ \frac{d \langle u'_i u'_j \rangle_{pat}}{dt} & = \left[ \frac{d \langle u'_i u'_j \rangle_{pat}}{dt} \right]_{sur} + \left[ \frac{d \langle u'_i u'_j \rangle_{pat}}{dt} \right]_{mix} + \left[ \frac{d \langle u'_i u'_j \rangle_{pat}}{dt} \right]_{bio}
\end{aligned} \label{eq:stat_model}
\end{equation}

Each term in the equations is developed with different methods following \cite{ser2023lagrangian}.
The entrainment terms are functions of the horizontal and vertical components of the entrainment rate $\Omega(t) = \frac{1}{V(t)} \frac{dV(t)}{dt}$ and of first and second moments of the tracer distributions:

\begin{equation}
 \left[ \frac{d \langle u_i \rangle_{pat}}{dt} \right]_{sur} = \Omega^h(t) \langle h_i \rangle_{suh} + \frac{1}{2} \Omega^v(t) \left( \langle a_i \rangle_{sua} + \langle b_i \rangle_{sub} \right) - \Omega(t) \langle u_i \rangle_{pat} \label{eq:entr_mean}
\end{equation}

\begin{equation}
\begin{aligned}
\left[ \frac{d \langle u'_i u'_j \rangle_{pat}}{dt} \right]_{sur} & = \Omega^h(t) \left( \langle h'_i h'_j \rangle_{suh} - \langle u'_i u'_j \rangle_{pat} + (\langle h_i \rangle_{suh} - \langle u_i \rangle_{pat})\cdot (\langle h_j \rangle_{suh} - \langle u_j \rangle_{pat}) \right) + \\ & + \frac{1}{2} \Omega^v(t) \left( \langle a'_i a'_j \rangle_{sua} - \langle u'_i u'_j \rangle_{pat} + (\langle a_i \rangle_{sua} - \langle u_i \rangle_{pat})\cdot (\langle a_j \rangle_{sua} - \langle u_j \rangle_{pat}) \right) + \\ & + \frac{1}{2} \Omega^v(t) \left( \langle b'_i b'_j \rangle_{sub} - \langle u'_i u'_j \rangle_{pat} + (\langle b_i \rangle_{sub} - \langle u_i \rangle_{pat})\cdot (\langle b_j \rangle_{sub} - \langle u_j \rangle_{pat}) \right) \\ 
\end{aligned} \label{eq:entr_cov_methods}
\end{equation}

The mixing term is involved in the size-dependent relaxation of the second moments:

\begin{equation}
\begin{aligned}
\left[ \frac{d \langle u'_i u'_j \rangle_{pat}}{dt} \right]_{mix} & = - \left( \frac{\kappa^h(t)}{(W(t))^2} + \frac{\kappa^v(t)}{(T(t))^2}\right)\langle u'_i u'_j \rangle_{pat} \\
\end{aligned} \label{eq:mix_cov}
\end{equation}

Finally, the interaction terms are the result of a Taylor expansion, whose details can be found in the Supplementary material.

\subsection{Biogeochemical modeling}

The biological interaction component (iii) of the model includes a single species of phytoplankton $P(\mathbf{x}, t)$, whose growth is regulated by a key limiting nutrient $N(\mathbf{x}, t)$ (Liebig's law \cite{tilman1982resource}) and its loss is parameterized as a linear function (other functions were tried but with small differences in the results). We express both distributions using nutrient's currency. The equations for the change in time of the two tracers are:

\begin{equation}
\begin{aligned}
\left[ \frac{d N(\mathbf{x}, t)}{dt} \right]_{bio} & = - \nu \frac{N(\mathbf{x}, t)}{N(\mathbf{x}, t)+k} P(\mathbf{x}, t) \equiv S_{upt}(\mathbf{x}, t) \\ \left[ \frac{d P(\mathbf{x}, t)}{dt} \right]_{bio} & =  + \nu \frac{N(\mathbf{x}, t)}{N(\mathbf{x}, t)+k} P(\mathbf{x}, t) - m P(\mathbf{x}, t) \\ 
\end{aligned} \label{eq:model_bio}
\end{equation}

where $\nu$ is the phytoplankton maximum growth rate, $k$ is its half saturation constant with respect to the limiting nutrient and $m$ is the phytoplankton mortality rate. 

The biogeochemical variation of DIC (dissolved inorganic carbon) is initially expressed as \cite{williams2011ocean}

\begin{equation}
\begin{aligned}
\left[ \frac{d DIC(\mathbf{x}, t)}{dt} \right]_{bio} & =  -\frac{K_o K_g \Delta p(\mathbf{x}, t)}{h} + \frac{1}{\rho} \frac{1}{R_{N:C}} S_{upt}(\mathbf{x}, t) 
\end{aligned} \label{eq:DIC_evol}
\end{equation}

The first term on the right-hand side is related to air-sea exchange of carbon dioxide, $K_o$ is the $CO_2$ solubility, $K_g$ is the average gas transfer velocity, $h$ is the mixed layer depth and $\Delta p$ is the difference in partial pressure of $CO_2$ between the water and the atmosphere.
The second term in instead related to biological carbon uptake (equation (\ref{eq:model_bio})), with $\rho$ that is the seawater density and $R_{N:C}$ the Redfield ratio between the nutrient $N$ of interest and carbon.

We can exploit the decomposition of the DIC in saturated and residual concentrations

\begin{equation}
    DIC(\mathbf{x}, t) = C_{sat}(\mathbf{x}, t) + \Delta C(\mathbf{x}, t)
\end{equation}

to introduce the Revelle buffer factor \cite{williams2011ocean}

\begin{equation}
    B = \left. \frac{\Delta p}{pCO_2} \right/ \frac{\Delta C}{C_{sat}} \simeq const.
\end{equation}

which can be assumed as constant for the ranges of $pCO_2 \sim pCO_2^{atm}$ considered. 
At this point, equation (\ref{eq:DIC_evol}) can be rewritten in terms of $\Delta p$, since

\begin{equation}
    \Delta C (\mathbf{x}, t) = \frac{C_{sat}(\mathbf{x}, t)}{B \cdot pCO_2^{atm}}  \Delta p(\mathbf{x}, t) \simeq \phi^{-1} \Delta p(\mathbf{x}, t)\label{eq:buf_res}
\end{equation}

and so 

\begin{equation}
\begin{aligned}
\left[ \frac{d \Delta p(\mathbf{x}, t)}{dt} \right]_{bio} & =   - \phi \frac{K_o K_g}{h} \Delta p(\mathbf{x}, t) + \frac{\phi}{\rho} \frac{1}{R_{N:C}} S_{upt}(\mathbf{x}, t) - \phi \Sigma_{\mathcal{T}} \frac{d \mathcal{T}(\mathbf{x}, t)}{d t} \\ 
\end{aligned} \label{eq:model_pco2_3}
\end{equation}

with the final assumption that saturated component's variation is mainly driven by changes in temperature $\mathcal{T}$.

\subsection{SOIREE simulation setup}

The biochemical model (Eq. (\ref{eq:model_bio})) relies on a Holling functional response of type II and a linear mortality rate, even if other setups were investigated.
The half-saturation constant $k$ is fixed to values typically observed for diatoms \cite{lancelot2000modeling, timmermans2001growth}, while the biological growth and mortality rates, $\nu$ and $m$, 
%are fitted to reproduce the experimental bloom magnitude.
are retrieved from SOIREE experimental data on the mean pyhtoplankton concentration.
In particular, we adopt a standard MSE minimization technique which estimates $\nu = 0.345\ d^{-1}$ and $m = 0.030\ d^{-1}$ in the physical-entrainment scenario, while $\nu = 0.407\ d^{-1}$ and $m = 0.055\ d^{-1}$ in the constant-entrainment case.
Carbon-related parameters used in equation (\ref{eq:model_pco2}) are selected from multiple sources \cite{li1971solubility, hannon2001modeling, williams2011ocean, morel1991iron, lancelot2000modeling} (see Table \ref{tab:soiree_param}).
The increase of temperature inside the patch in the first two weeks is assumed to be linear and of magnitude $\simeq 0.03$ \textdegree$\mathrm{C/day}$.
During SOIREE, in situ casts reported only a modest decrease in the mixed layer depth. For this reason, and considering the low values of $\delta_0$ and $\kappa^v_0$ characteristic of the region (Table \ref{tab:soiree_param}), no significant change in thickness is obtained.

Initial conditions of the water patch and of the tracer concentrations are set to mimic the environment where the iron release took place \cite{abraham2000importance, hannon2001modeling, frew2001macronutrient}. 
Inside the patch, we impose $\langle N \rangle_{pat}(0) = 0.08\ \mathrm{(\mu mol\ Fe) \cdot m^{-3}}$ and $\langle P \rangle_{pat}(0) = 0.05\ \mathrm{(\mu mol\ Fe) \cdot m^{-3}}$ \cite{boyd2000mesoscale, abraham2000importance}, which are the same values we adopt also for the horizontal and above surrounding (see Supplementary).
%\commentos{below the mixed layer from data on chl there is more chl, with P = 0.07}.
The different injections made in the first two weeks of the experiment are not explicitly characterized, 
but the dissolved iron mean concentration is fixed to what is observed after each release \cite{bcodmofesur}, \cite{bcodmofever}, being free to vary in the intermediate periods (Fig. \ref{fig:N-P_simple}C). 
At day $8$ of the experiment, dissolved iron reaches a concentration which is about $23$ times higher than the horizontal surrounding and $6$ times higher than the one below.
From there, the decay of iron is modelled only through the biogeochemical dynamic we have taken into consideration. 
Regarding carbon and its composition, we made the assumption that the mean $pCO_2$ difference ($\langle \Delta p \rangle$) is initially at the equilibrium of equation (\ref{eq:model_pco2}) at time $t=0$ both inside and in the outside regions. 
Contrary to the mean surrounding concentrations of iron and phytoplankton that we approximate as constant \cite{bcodmofesur, bcodmofever, bcodmochl}, we choose to resolve the evolution in time of $\langle \Delta p^h \rangle_{suh}$ and $\langle \Delta p^a \rangle_{sua}$ using always the equation (\ref{eq:model_pco2}) but integrating in different regions of space.
Finally, since no enough clear information on the initial concentration variance and covariance of the other tracers exists, we fix them to zero, assuming initial homogeneity.

\begin{table} [h]
    \centering
    \begin{tabular}{@{} *4l @{}} \toprule
         \textbf{parameter}& \textbf{value}&\textbf{unit of measure}&\textbf{source}\\\midrule
         $\gamma_0$& $0.12$&$\mathrm{day^{-1}}$&\cite{ser2023lagrangian}\\ 
         $\delta_0$& $2.88\cdot 10^{-3}$&$\mathrm{day^{-1}}$&\cite{hannon2001modeling}\\
         $\kappa^h_0$& $0.10$&$\mathrm{km^2 \cdot day^{-1}}$&\cite{ser2023lagrangian}\\
         $\kappa^v_0$& $2.07\cdot 10^{-6}$&$\mathrm{km^2 \cdot day^{-1}}$&\cite{hannon2001modeling}\\ 
         $k$& $1.20$&$\mathrm{(\mu mol\ Fe) \cdot m^{-3}}$&\cite{lancelot2000modeling}\\
         $\phi$&$2.29$&$\mathrm{\mu atm \cdot kg \cdot (\mu mol\ C)^{-1}}$&\cite{hannon2001modeling}\\
         $K_o$&$0.06$&$\mathrm{(\mu mol\ C) \cdot kg^{-1} \cdot  \mu atm^{-1}} $&\cite{li1971solubility}\\
         $K_g$&$5.96$&$\mathrm{m \cdot day^{-1}}$&\cite{wanninkhof1992relationship}\\
         $h$&$65$&$\mathrm{m}$&\cite{hannon2001modeling}\\
         $\rho$&$1027$&$\mathrm{kg \cdot m^{-3}}$&\cite{boyd2001southern}\\
         $B$&$ 14 $ &$\mathrm{\#}$&\cite{bakker2001southern}\\
         $R_{Fe:C}$&$5 \cdot 10^{-5}$ &$\mathrm{(\mu mol\ Fe) \cdot (\mu mol\ C)^{-1}}$&\cite{bozec2005co2}\\
        $\Sigma_{\mathcal{T}}$&$ -8$&$\mathrm{(\mu mol\ C) \cdot kg^{-1} \cdot  K^{-1}} $&\cite{williams2011ocean}\\
        $R_{C:Chl}$&$60$ &$\mathrm{(\mu mol\ C) \cdot (\mu mol\ Chl)^{-1}}$&\cite{boyd2000mesoscale}\\\bottomrule
    \end{tabular}
    \caption{Set of parameters used to simulate the SOIREE iron fertilization experiment}
    \label{tab:soiree_param}
\end{table}

\subsection*{Data availability}
All external data used, such as observations or laboratory measures, are freely available from the corresponding publications cited along the text.

\subsection*{Competing interests}
The authors declare no competing interests.

\subsection*{Author contributions}
S.C. performed analytical \& numerical calculations and wrote the manuscript. All authors interpreted the results, suggested analysis and contributed in shaping the manuscript. E.S-G. directed the study.

\subsection*{Funding}
E.S.-G., S.D. and M.J.F. are very grateful for support from the Simons Foundation: the Simons
Collaboration on Computational BIOgeochemical modeling of Marine EcosystemS (CBIOMES \#549931 to M.J.F.). E.S.G acknowledges María de Maeztu Project No. CEX2021-001164-M, funded by the MCIN/AEI/10.13039/501100011033 and funding from the Agencia Estatal de Investigaciòn (MICIU/AEI/10.13039/501100011033) and FEDER “Una manera de hacer Europa” under Project LAMARCA No. PID2021-123352OB-C32. S.C. and E.S.G acknowledge the Beatriz Galindo Fellowship (BG22-00038), funded by the Spanish Ministry of Universities.

\subsection*{Ethics declaration}
Not applicable

\printbibliography

\end{document}